\documentstyle[preprint,prl,aps]{revtex}
\draft

\begin{document}
\title{{\boldmath $C\!P$} violation in B mesons using Dalitz plot 
asymmetries}
\author{Rahul Sinha
\ifpreprintsty
\footnote{e-mail: sinha@imsc.ernet.in}
\else
\cite{author}
\fi
}
\address{ Institute of Mathematical Sciences, Taramani, Madras 600113, 
India.}
\date{\today}
\maketitle
\begin{abstract}            
\widetext
We study $C\!P$~violation in $B\to K^* \ell^+\ell^-$  using generalized 
Dalitz plot asymmetries in the angular distribution.  These new kind of 
asymmetries are constructed by adding $B$ and ${\bar B}$  events, and 
do not require flavor or time tagging,  nor is the presence of strong 
phases needed. Using this method one requires about $2 \times 10^8/\eta$ 
$B$'s to measure the $C\!P$~violating parameter $\eta$ in the Standard 
model. The two--Higgs doublet model requires only $10^7$ $B$'s to 
constrain parameters better than done by the electric dipole moment of 
the neutron.
\narrowtext
\end{abstract}
 
$B$ mesons are expected to exhibit $C\!P$~violation like the $K$~
mesons, which is the only system where this phenomena has been observed. 
Much effort has thus been devoted to studying possible signals of 
$C\!P$~violation in $B$. Large numbers of $B$ mesons are expected to be 
produced in the future, which would enable study of its rare decay modes. 
However, since flavor and time tagging is difficult, except at the 
asymmetric $e^+e^-$ factories, the large number of $B$'s cannot be 
efficiently used to study $C\!P$~violation. 
 
In view of this difficulty Burdman and Donoghue\cite{Burdman} studied 
the possibility of detecting $C\!P$~violation in $B$ decays without the 
need for flavor identification. Much like the Dalitz plot 
asymmetry\cite{Okubo} for $K^\pm$, Ref.\cite{Burdman} considers 
asymmetries in the hadronic three body decays of neutral $B$ mesons to 
flavor states that are $C$ eigenstates, as well as to states which under 
$C$ conjugation return to $C$ partners. A very important point 
realized by these authors is that if {\em ``one searches for quantities for 
which $C\!P$~invariance says that they should change sign when 
comparing $B$ and ${\bar B}$ decays, summing $B$ and ${\bar B}$ should 
produce a net null result unless $C\!P$~is violated.''} Since such 
techniques do not involve flavor identification, they do not depend on 
the production characteristics and can be studied using any source of 
$B$ mesons. These Dalitz plot asymmetries are logically distinct from 
the partial rate asymmetries usually considered, in the sense that they 
may be present even when partial rate asymmetries vanish. 

In this letter we discuss such asymmetries in angular variables that 
require no flavor or time tagging and also {\em do not need strong 
phases to show up}. These asymmetries are independent of CKM phase of 
$B-{\bar B}$ mixing, and depend only on direct or indirect CP violation.  
We construct such asymmetries for the rare decay $B\to K^* \ell^+\ell^-$ 
($\ell^+ \ell^-$ non resonant), where new physics contributions are 
expected to show up. Our choice of mode is such that the asymmetries are 
free from undetermined strong phases. Another valuable feature is a 
clean signal which will prove to be one of the easiest to measure. Our 
analysis is similar  in spirit to the approach of Ref.\cite{Sehgal} 
where CP violation in $K_L\to \pi\pi e^+e^-$ has been considered.

The  effective short distance Hamiltonian relevant to the decay $b\to s 
\ell^+\ell^-$ \cite{Buras-Munz,Grinstein,ODonnell} leads to the QCD 
corrected matrix element 
\widetext
\begin{equation}
{\cal M}
(b\to s \ell^+\ell^-)
=\frac{\alpha G_F}{\sqrt{2}\pi} \sum_{j} v_j \{
-2 i C_7^j m_b \frac{q^\nu}{q^2}\;\bar{s}\sigma_{\mu\nu}b_R\; 
\bar{\ell}\gamma^\mu\ell 
+C_8^j\;\bar{s}\gamma_\mu b_L\;\bar{\ell}\gamma^\mu\ell 
+C_9^j\;\bar{s}\gamma_\mu b_L\;\bar{\ell}\gamma^\mu\gamma_5\ell 
\}
\label{heff}
\end{equation}
\narrowtext
\noindent
where the sum j is over all the flavors ``$u,c,t$'' in the loops, 
$C_{7,8,9}^j$ are the Wilson coefficients given in 
Ref.\cite{Buras-Munz,Grinstein}, $m_b$ is the mass of the b quark, $q^2$  
is the invariant lepton mass squared, $b_{L,R}=(1\mp\gamma_5)/2\,b$ and 
$v_j=V^*_{js}V_{jb}$ is the product of the CKM matrix elements.

The transition matrix element for the exclusive process $B(p)\to K^*(k) 
\ell^+\ell^-\to K(k_1) \pi (k_2)\ell^+(q_1)\ell^-(q_2)$ can be written 
for each of the operators in Eq.(\ref{heff}) as,
\[
\langle K\pi|\bar{s}i\sigma_{\mu\nu}(1\pm\gamma_5)q^\nu b|B\rangle = 
i{\cal A} \epsilon_{\mu\nu\alpha\beta} K^\nu k^\alpha q^\beta\pm {\cal 
B}K_\mu\pm {\cal C} k_\mu \,,
\]
\[
\langle K\pi|\bar{s}\gamma_\mu(1\mp\gamma_5)b|B\rangle = i{\cal 
D}\epsilon_{\mu\nu\alpha\beta} 
K^\nu k^\alpha q^\beta \pm {\cal E} K_\mu\pm {\cal F} k_\mu \,.
\] 
The form-factors ${\cal A,\cdots,F}$ are unknown functions of 
$q^2=(p-k)^2$ and other dot products involving momentum, $k=k_1+k_2$ and 
$K=k_1-k_2$ and can be related to those used in \cite{Wyler}, 
\cite{Burdman_new} as given in Table~\ref{tab1}. The variable $\sigma$ 
arises due to the decay of $K^*\to K\pi$, evaluated in the zero width 
approximation. The  current proportional to $q_\mu$ does not contribute 
as it couples to light leptons. In our notation $M_B,m_{K^*},m_K$ and 
$m_\pi$ are the masses of the $B$, $K^*$, $K$ mesons and the pion 
respectively. We first consider the case of charged B's and later 
generalize to neutral ones. The matrix element for the process $B\to K^* 
\ell^+\ell^-\to K \pi \ell^+\ell^-$  can be written as
\ifpreprintsty
\[
{\cal M}( B \displaystyle\to
K\pi\ell^+\ell^-)=\frac{\alpha G_F}{\sqrt{2}\pi}
\left\{\left(i \alpha_L\,\epsilon_{\mu\nu\alpha\beta}\, K^\nu k^\alpha 
q^\beta\right.\right. 
  \displaystyle\left. \left.+ \beta_L\, K_\mu+ \rho_L\, k_\mu
\right){\bar\ell}\gamma_\mu\,L\,\ell+ L\to R
\right\}\,,
\]
\else
\begin{eqnarray}
{\cal M}( B &\displaystyle\to&
K\pi\ell^+\ell^-)=\frac{\alpha G_F}{\sqrt{2}\pi}
\left\{\left(i \alpha_L\,\epsilon_{\mu\nu\alpha\beta}\, K^\nu k^\alpha 
q^\beta\right.\right.
\nonumber \\[2ex] 
  & &\displaystyle\left. \left.+ \beta_L\, K_\mu+ \rho_L\, k_\mu
\right){\bar\ell}\gamma_\mu\,L\,\ell+ L\to R
\right\}\,,
        \nonumber
\end{eqnarray}
\fi
where $L,R=\displaystyle\frac{(1\mp\gamma_5)}{2}$, $q=q_1+q_2$ and 
$Q=q_1-q_2$, and the coefficients $\alpha_{L,R},~\beta_{L,R}$ and 
$\rho_{L,R}$ are given by
\widetext
\begin{equation}
\begin{array}{rclcrcl}
\displaystyle\alpha_{R,L} &=&\displaystyle 
\sum_j\,v_j\,\{\frac{(C_8^j\pm C_9^j)}{2}\,{\cal D}
       -\frac{m_b}{q^2}\,C_7^j\,{\cal A}\}&=&\displaystyle\sum_j|{\rm
a^j}_{R,L}|\,exp\left(i\delta^{\alpha_j}_{R,L}\right)\,exp 
\left(i\phi^{\alpha_j}_{R,L}\right) 
\\[2ex] 
\displaystyle\beta_{R,L}& =&\displaystyle \sum_j\,v_j\,\{\frac{(C_8^j\pm 
C_9^j)}{2}\,{\cal E}
       -\frac{m_b}{q^2}\,C_7^j\,{\cal B}\}&=&\displaystyle\sum_j|{\rm
b^j}_{R,L}|\,exp\left(i\delta^{\beta_j}_{R,L}\right)\,
exp\left(i\phi^{\beta_j}_{R,L}\right) 
\\[2ex]    
\displaystyle\rho_{R,L} &=&\displaystyle \sum_j\,v_j\,\{\frac{(C_8^j\pm 
C_9^j)}{2}\,{\cal F} 
       -\frac{m_b}{q^2}\,C_7^j\,{\cal C}\}&=&\displaystyle\sum_j|{\rm
r^j}_{R,L}|\,exp\left(i\delta^{\rho_j}_{R,L}\right)\,
exp\left(i\phi^{\rho_j}_{R,L}\right) \,.
\end{array}
	\label{alpha_beta_rho}
\end{equation} 
\narrowtext
\noindent
In the above equation $\alpha$, $\beta$ and $\rho$ are recast in terms 
of $a$, $b$ and $r$ so as to identify the strong phases $\delta$ and the 
weak phases $\phi$. Using $CPT$  invariance, the matrix element for the 
decay ${\bar B}\to\bar K \bar\pi\ell^+\ell^-$ can be obtained from the 
$B\to K\pi\ell^+\ell^-$ by replacing $\alpha_{L,R}\to 
-{\bar\alpha}_{L,R}, \beta_{L,R}\to {\bar\beta}_{L,R}, 
\rho_{L,R}\to{\bar\rho}_{L,R}$ \cite{valencia,okubo2}, where  
\begin{equation}
\displaystyle{\bar \alpha}_{R,L}=\displaystyle\sum_j|{\rm
a^j}_{R,L}|\,exp\left(i\delta^{\alpha_j}_{R,L}\right)\, 
exp\left(-i\phi^{\alpha_j}_{R,L}\right) 
	\label{alpha_beta_rho_bar}
\end{equation}

and similar relations hold for $\bar\beta$ and $\bar\rho$. The matrix 
element mod. squared for the process $B\to K^* \ell^+\ell^-\to K \pi 
\ell^+\ell^-$  is worked out retaining the imaginary parts in 
$\alpha,\beta$ and $\rho$ and presented in Table~\ref{tab2}. We define 
$X$ as the three momentum of the $\ell^+\ell^-$ or $K\pi$ invariant 
system in the B meson rest frame and $\lambda_K(\lambda_e)$ is related 
to the $K(e)$ three momentum in the $K^*(e^+e^-)$ rest frame . 
$\theta_e(\theta_K)$ is the angle of the $e^-(K)$ in the $e^+e^-(K\pi)$ 
rest frame with the $e^+e^-(K^*)$ invariant direction. $\varphi$ is the 
angle between the planes defined by $e^+e^-$  and the $K\pi$ directions. 
Our choice of variables and the general treatment presented so far 
resembles the formalism developed for the $K_{\ell 4}$ 
decays\cite{pias_treiman}. The essential difference is, that while the 
latter was aimed at obtaining the $\pi-\pi$ phase shifts  {\it i.e.}  
strong phases, our interest here is in the $C\!P$~violating weak 
phases. The differential decay rate is then given by
\[
d\Gamma=\displaystyle\frac{1}{2^{14} \pi^6 M_B^2}\int|{\cal M}|^2 
X\lambda_K \lambda_e\, dq^2 d\cos\theta_K d\cos\theta_e d\varphi\,,
\]
assuming a narrow width approximation for the decay $K^*\to K\pi$. 

It can easily be seen from Table~\ref{tab2} that, the only terms 
proportional  to $\sin(\varphi)$ or $\sin(2 \varphi)$ are those  that 
depend on the imaginary parts of the $\alpha,~\beta$ or $\rho$. For 
instance only the coefficient of ${\rm 
Im}(\alpha_L\beta^*_L+\alpha_R\beta^*_R)$ is proportional to $\sin(2 
\varphi)$. Hence we can isolate this term by considering the following 
asymmetry in terms of the differential decay rates of the B meson with 
respect to $\varphi$,
\begin{equation}
\displaystyle A_1=\displaystyle \frac{1}{\Gamma}
 (\displaystyle\int_0^{\frac{\pi}{2}}
-\int_{\frac{\pi}{2}}^\pi
+\int_\pi^{\frac{3\pi}{2}}
-\int_{\frac{3\pi}{2}}^{2\pi})\frac{d\Gamma}{d\varphi}d\varphi\,.
\end{equation}
The imaginary part in the term under consideration can be due to either 
a strong phase or a weak phase. An astute reader will, however, have 
realized that such $C\!P$~violating asymmetries can be obtained not by 
considering the difference of differential rates for $B$ and ${\bar B}$ 
, but the sum of these rates. It follows trivially from 
eqn.(\ref{alpha_beta_rho} and \ref{alpha_beta_rho_bar}) that the 
asymmetry for $B$(${\bar B}$ ) is
\[
A_1({\bar A_1})   \propto\pm\displaystyle \sum_{j,k} \{\displaystyle 
|a^j_L| |b^k_L| \sin(\displaystyle(\delta_L^{jk})\pm(\phi_L^{jk}))+L \to 
R \}
\]
where $\delta_L^{jk}\equiv(\delta_L^{\alpha_j}-\delta_L^{\beta_k})$ and 
$\phi_L^{jk}\equiv(\phi_L^{\alpha_j}-\delta_L^{\beta_k})$. The sum of 
the two asymmetries ${\sf A}_1^{CP}=  A_1+{\bar A_1}$ becomes
\begin{equation} 
{\sf A}_1^{CP}\propto\displaystyle \sum_{j,k} \{\displaystyle |a^j_L| 
|b^k_L| \cos(\displaystyle\delta_L^{jk})\,\sin(\phi_L^{jk})+L \to R \} 
\,,\label{asymm} 
\end{equation} 
which is {\em nonzero if and only if there is $C\!P$~violation 
represented by non-zero phases $\phi$}\cite{Burdman,valencia}. For $B\to 
K^*\ell^+\ell^-$, $\delta$ can arise either from the quark in the 
penguin loop going on shell, {\it i.e.} $q^2\geq 4\,m_q^2$, or from 
electromagnetic final state interactions which are negligible and 
ignored. The case of quark on shell is taken care of in evaluating the 
coefficients $C^j_{7,8,9}$ and included in our analysis. It is also 
possible to construct a different asymmetry that isolates another 
combination of the imaginary terms. Such  an asymmetry 
\cite{kramer-palmer} considers the difference distribution of the same 
hemisphere and opposite hemisphere events, and can be defined by
\begin{equation}  
\displaystyle A_2= \displaystyle 
 \frac{1}{\Gamma}(\displaystyle\int_0^\pi-\int_\pi^{2\pi}) 
 d\varphi\displaystyle\int_Dd\cos\theta_e\int_Dd\cos\theta_K 
 \tilde{\Gamma}
\end{equation}
where $\displaystyle\int_D\equiv\displaystyle\int_{-1}^0-\int_0^{1}$ and 
$\tilde{\Gamma}=\displaystyle\frac{d\Gamma}{d\cos\theta_e d\cos\theta_K 
d\varphi}$.

        For neutral $B$ mesons the asymmetries are even more 
interesting, as it is here, that flavor tagging not being needed, is a 
real advantage. In addition we also find that the time integrated 
asymmetries are independent of the parameters describing the 
oscillation, and  reduce to the asymmetries for charged $B$'s. The time 
evolution of $B$ mesons is given by  
\begin{eqnarray}
\displaystyle|B^0(t)\rangle &=&\displaystyle 
g_{+}(t)|B^0\rangle+\frac{1}{\xi} g_{-}(t)|{\bar B^0} \rangle \\ 
\nonumber
\displaystyle|{\bar B^0}(t)\rangle &=&\displaystyle g_{+}(t)|{\bar B^0} 
\rangle+\xi g_{-}(t)|B^0\rangle  
\end{eqnarray}
with $ \displaystyle g_{\pm}=\displaystyle \exp\{\displaystyle 
-(\frac{\Gamma_B}{2}-i\,M_B)t\} (\cos\displaystyle\frac{\Delta M\, 
t}{2},i \sin\displaystyle\frac{\Delta M\, t}{2}) $ and  
$\xi=\displaystyle\frac{p}{q}$. $M_B(\Gamma_B)$ and $\Delta 
M(\Delta\Gamma)$ are the average and the difference of the masses 
(widths) of the two mass eigenstates $B_H$ and $B_L$ respectively. Hence 
for $B^0$ mesons we have  
\ifpreprintsty
\[
{\cal M}( B^0(t) \displaystyle \to
K\pi\ell^+\ell^-)=\frac{\alpha G_F}{\sqrt{2}\pi}
\displaystyle\left\{\left(i {\alpha^0_L}\,\epsilon_{\mu\nu\alpha\beta}\, 
K^\nu k^\alpha q^\beta\right.\right.
 \displaystyle\left. \left.  +{ \beta^0_L}\, K_\mu+ {\rho^0_L}\, k_\mu
\right){\bar\ell}\gamma_\mu\,L\,\ell+L\to R \right\}\,,
\]
\else
\begin{eqnarray}
{\cal M}( B^0(t) &\displaystyle \to&
K\pi\ell^+\ell^-)=\frac{\alpha G_F}{\sqrt{2}\pi}
\displaystyle\left\{\left(i {\alpha^0_L}\,\epsilon_{\mu\nu\alpha\beta}\, 
K^\nu k^\alpha q^\beta\right.\right.
\nonumber\\[2ex]
  & & \displaystyle\left. \left.  +{ \beta^0_L}\, K_\mu+ {\rho^0_L}\, 
  k_\mu 
\right){\bar\ell}\gamma_\mu\,L\,\ell+L\to R \right\}\,, \nonumber
\end{eqnarray}
\fi
with an analogous relation for $\bar B^0\to \bar K\bar \pi\ell^+\ell^-$ 
written by replacing $\alpha^0_{L,R}\to 
{\bar\alpha^0_{L,R}},\beta^0_{L,R}\to {\bar\beta^0_{L,R}}, 
\rho^0_{L,R}\to {\bar\rho^0_{L,R}}$, where 
\begin{equation}
\begin{array}{rlrl}
\displaystyle \alpha^0=&\displaystyle g_{+}\;\alpha-\xi^{-1}\, g_{-}\; 
{\bar \alpha}\,, &
\displaystyle\overline{\alpha^0}=&\displaystyle \xi\, g_{-}\;\alpha- 
g_{+}\; {\bar \alpha} \\                        
\displaystyle \beta^0=&\displaystyle g_{+}\;\beta+\xi^{-1}\, g_{-}\;
{\bar \beta}\,, &
\displaystyle\overline
{\beta^0}=&\displaystyle \xi\, g_{-}\;\beta+
g_{+}\;{\bar \beta} \\
\displaystyle \rho^0=&\displaystyle g_{+}\;\rho+\xi^{-1}\, g_{-}\;
{\bar \rho}\,, &
\displaystyle\overline{\rho^0}=&\displaystyle \xi\, g_{-}\;\rho+
g_{+}\; {\bar \rho}\,. \\
\end{array}
\end{equation}
Here we have suppressed the subscripts $L,R$, and a summation over both 
is implied, in what follows.

By adding differential decay rates for $B^0$ and $\bar{B^0}$, one can 
construct an asymmetry of the type ${\sf A}_1^{CP}$, which shall be 
proportional to ${\alpha}^0 {\beta^0}^*+\overline{{\alpha}^0}\overline{ 
{\beta^0}^*}$, and given by
\widetext
\begin{equation}
\begin{array}{rl}
\displaystyle{{\sf A}_1^0}^{CP}(t)\propto &\displaystyle e^{-\Gamma_B 
t}\left\{ {\rm Im}\left(\alpha\,\beta^*-\bar\alpha 
\bar\beta^*\right)\,\cos^2\frac{\Delta M t}{2}+\displaystyle{\rm 
Im}\left( \left|\displaystyle\xi\right|^2 \alpha\,\beta^*- 
\left|\displaystyle\xi^{-1}\right|^2 \bar\alpha 
\bar\beta^*\right)\,\sin^2\frac{\Delta M t}{2}\right.\\
-&\left.\displaystyle\,\frac{i}{2}{\rm Im}\left(
\left[\xi^*-\xi^{-1}\right]\bar\alpha\beta^*+
\left[\xi-(\xi^*)^{-1}\right]\alpha\bar\beta^*\right) 
\sin(\Delta M t)\right\}\,.
\end{array}
\end{equation}
\narrowtext
For $\displaystyle\xi=e^{2 i \beta}$ one can easily see that this 
asymmetry has the form similar to eqn.(\ref{asymm}), and indeed is the 
same after time integration. One should note that no $\beta$ dependence 
survives in the asymmetry, where $\beta$ is the CKM phase of the 
$B-{\bar B}$ mixing diagram. Hence, the asymmetries under consideration 
in $B\to K^* \ell^+\ell^-$  are insensitive to mixing induced $C\!P$~
violation and measure only direct $C\!P$~violation. However, if 
$\displaystyle\xi=\displaystyle \frac{1+\epsilon} {1-\epsilon}$ the 
asymmetry involves both  ${\rm Re}(\epsilon)$ and the phases $\phi$:
\widetext
\ifpreprintsty
\begin{eqnarray}
\displaystyle{{\sf A}_1^0}^{CP}(t)\propto \sum_{j,k}
&&\displaystyle
e^{-\Gamma_B t}|a^j||b^k|\left\{ 2 \cos(\displaystyle\delta^{jk})\, 
\sin(\phi^{jk}) \right. +\displaystyle 8\,{\rm 
Re}(\epsilon)\;\sin(\displaystyle\delta^{jk})\, 
\cos(\phi^{jk})\;\sin^2\frac{\Delta M t}{2}\nonumber \\
+&&\displaystyle\left. 4\,{\rm 
Re}(\epsilon)\;\cos(\displaystyle\delta^{jk})\,\cos(\phi^{jk})\;
\sin(\Delta M t)\right\}\,.\nonumber
\end{eqnarray}
\else
\[                                  
\displaystyle{
{\sf A}_1^0}^{CP}\!(t)\!\propto \!
\displaystyle
e^{-\Gamma_B t}|a^j||b^k|\left\{\cos(\displaystyle\delta^{jk})
\sin(\phi^{jk})
\right.
+\displaystyle 4{\rm Re}(\epsilon)\;\sin(\displaystyle\delta^{jk}) 
\cos(\phi^{jk})\;\sin^2\frac{\Delta M t}{2} +\displaystyle\left. 2{\rm 
Re}(\epsilon)\;\cos(\displaystyle \delta^{jk})\cos(\phi^{jk})\; 
\sin(\Delta M t)\right\}.
\]
\fi
\narrowtext
For the $B_d$ system, ${\rm Re}(\epsilon)$ is expected to be $10^{-3}$. 
The measured upper limit is ${\rm Re}(\epsilon)\textstyle< 
0.045$ at 90\% C.L.\cite{pdg}.  Hence in estimating the number of $B_d$ mesons 
required to detect $C\!P$~violation we assume ${\rm Re}(\epsilon)=0$.

$C\!P$~violating asymmetries in $b \to (s,d)\gamma\,,(s,d)
\ell^+\ell^-$ have been considered\cite{Soares,Wolfenstien,chinese}
recently, for both Standard model (SM) as well as the two--Higgs
doublet model (2HDM). However all these discussions
require  flavor tagging and in most cases  rely on the presence of large 
strong phases arising out of final state interactions. Since very large 
numbers of $B$'s are required to detect $C\!P$~violation in these modes 
(several hundreds of times more than that possible at asymmetric 
$B$-factories), a technique that does not require flavor or time tagging 
would clearly be beneficial. We construct $C\!P$~violating asymmetries 
of the type ${\sf A}_1^{CP}$ and ${\sf A}_2^{CP}$ for SM as well as 
2HDM.  Here we present only an estimate of the number of B mesons 
required to observe a $C\!P$~violating asymmetry. We choose to present 
our results in this form rather than numbers for asymmetries, so as to 
minimize the dependence of our results on non CP violating parameters of 
the models. Details of our numerical work will be presented elsewhere. 
In estimating numbers we have ignored statistical and systematic errors, 
which will be a part of any experiment measuring such asymmetries. The 
experimental procedure to observe $C\!P$~violation will assume that 
there is a sample with {\em equal numbers} of $B$ and ${\bar B}$. It 
would be imperative that the cuts imposed are such as to minimize 
inherent asymmetry in the collected samples of $B$ and ${\bar B}$. We 
use the form factors from the quark model (QM) of Ref.\cite{Wyler}, 
since in heavy quark effective theory (HQET) they cannot currently  be 
reliably predicted over the entire dilepton mass range. However, use of HQET  
will be possible once $B\to\rho\ell\nu$ data is available. At the time 
these asymmetries are experimentally studied this data should presumably 
be available and HQET would be choice for form factors. It has recently 
been shown \cite{deshpande} that contrary to held beliefs, 
if an interplay of weak and strong phases of two different amplitudes in 
addition to $B^0-\overline{B^0}$ mixing are considered, it is possible 
to detect $C\!P$~violation at symmetric $e^+e^-$ colliders. {\em We 
have shown that this is possible even without the presence of strong 
phases}. All arguments presented here are equally applicable to the 
processes $B\to \rho \ell^+\ell^-\to \pi \pi \ell^+\ell^-$ and $B_s\to 
\phi \ell^+\ell^-\to K K \ell^+\ell^-$ . 

\widetext
In  SM the $C\!P$~violating asymmetries ${\sf A}_1^{CP}$ and ${\sf
A}_2^{CP}$  take the form  
\begin{equation}
{\sf A}_1^{CP}=\displaystyle 2\,x\,A^2 \lambda^6\eta\,\Delta\,\int\,dq^2 
\,{\sf C}\, ({a_0\,V-A_0\,g})\,X^2\, ,\;
{\sf A}_2^{CP}=\displaystyle\,x\,A^2 \lambda^6\eta\,\int\,dq^2 \,{\sf 
C}\,{\sf F} \frac{1}{\sqrt{q^2}}\,X^2\;.
\end{equation}
where
$x=\displaystyle\frac{\alpha^2 G_F^2\,\sigma^2\,m_b\,m_{K^*}\, 
\lambda^3_K }{2^{10} \,9 \pi^8 M_B (\Gamma+\bar\Gamma)(M_B+m_{K^*})}$, 
${\sf C}={\rm Re}(
{C_7^t}\,{C_8^c}^* - {C_7^c}^*\,{C_8^t}
+{C_7^u}\,{C_8^t}^*- {C_7^t}^*\,{C_8^u}
+{C_7^c}\,{C_8^u}^* - {C_7^u}^*\,{C_8^c})$,
and ${\sf F}=\left\{2 X^2 M_B^2 ({a_{+}\,V-A_{+}\,g})+
({a_0\,V-A_0\,g})\,\Delta\,{k\cdot q}\right\}$.
\narrowtext
\noindent
We estimate the number of $B$ mesons required to see $C\!P$~violation 
as ${\sf N}_1^{CP}\equiv ({\sf A}_1^{CP} \tau_B\,\Gamma)^{-1}\approx 
1\times10^{10}$ and ${\sf N}_2^{CP}\approx 2\times10^{8}$. Clearly 
asymmetry ${\sf A}_2^{CP}$ is more sensitive as should be expected.
  
In 2HDM CP violation arises from a relative phase between the two Higgs 
vacuum expectation values. Even if $\eta=0$ there is $C\!P$~violation 
in 2HDM. Only contribution from the top intermediate state in the loop 
is enough to generate $C\!P$~violation. For 2HDM each of the 
coefficients \cite{Grinstein} $C_7,~C_8$ and $C_9$ get extra contribution 
above those in the standard model.  However, only $C_7$ gets 
contribution from a complex phase, arising from the Higgs vacuum 
expectation values, giving
\[
C_7=\displaystyle C_7^{SM}+|\xi_t|^2\,\tilde{C_7^H}+{\rm 
Re}(\xi_t\xi_b)\,C_7^H+i\,{\rm Im}(\xi_t\xi_b)\,C_7^H\;.
\] 
The resultant asymmetries being given by equations identical to SM 
except that ${\sf C}={\rm Re}\!\left({C_7^t}\,{C_8^t}^*\right)$ and 
$\lambda^2 \eta \to {\rm Im}(\xi_t \xi_b)$. Using the constraint from 
electric dipole moment ({\it e.d.m.}) of the neutron, the upper limits
on ${\rm Im}(\xi_t \xi_b)$ 
range from 0.3 --10~\cite {Wolfenstien}  due to large uncertainities
in the hadronic form factors. 
We refer the reader to Ref.\cite {Wolfenstien} and refrain from details 
here. It is found that using ${\sf A}_2^{CP}$ we need $\approx 7\times 
10^6$ $B$'s to constrain $C\!P$~violating parameters of the 2HDM better 
than done by {\it e.d.m.} of the neutron.

To conclude we have studied $C\!P$~violating Dalitz plot asymmetries in 
angular variables. We consider such asymmetries for the process $B\to 
K^* \ell^+\ell^-\to K \pi \ell^+\ell^-$ .  These asymmetries are 
constructed by adding $B$ and ${\bar B}$ events, and do not require 
flavor or time tagging, nor do they depend on unknown strong phases.  
These asymmetries provide a clean signal for $C\!P$~violation that will 
prove easy to measure. Several experiments\cite{pdg} have already 
measured some angular distributions in a related process $B\to 
K^*J/\psi$.  $B\to K^* \ell^+\ell^-$  is likely to be seen soon, and one 
should either be able to see $C\!P$~violation or establish better 
better bounds on models of $C\!P$~violation like the 2HDM. Using the 
methods discussed here it should be possible to measure $\eta$ in 
planned future colliders.

I am grateful to Prof. G. Rajasekaran and Dr. N. Sinha for discussions 
and encouragement. I also thank Profs. R. Anishetty, S. R. Choudhury, R. 
H. Dalitz, H. S. Mani, S. Okubo and R. Ramachandran for discussions.

\begin{center}
\ifpreprintsty
\begin{table}
\else
\begin{table}
\squeezetable
\fi
\caption{ Relations between the form factors used in this paper, a quark 
model (QM) that reproduces heavy quark limit and heavy quark effective 
theory (HQET). $\displaystyle W_\mu=(K_\mu-\zeta\,k_\mu)$, $\sigma^2=96 
\pi^2/(m_{K^*}^2 {\lambda_K}^3)$, $\lambda_K,~\Delta$ and $\zeta$ are 
defined in Table~\protect\ref{tab2}.
}
\label{tab1}
\begin{tabular}{c|c|c}
        & QM \cite{Wyler} & HQET\cite{Burdman_new}\\
\hline
${\cal A}$ & $-2\,g\,\sigma$    & $(A+B)\,\sigma$ \\[0ex]
${\cal B}$ & $a_0\,\Delta\,\sigma$  & $\displaystyle
-(\frac{A+B}{2}\,\Delta+\frac{A-B}{2}\,q^2)\,\sigma $ \\[2ex] 
${\cal C}$ & $\displaystyle 2\,a_{+}\,W\cdot q\,\sigma-\zeta\,{\cal B}$ & 
  $\displaystyle 2\,W\cdot q\,(\frac{A+B}{2}+\frac{C}{2}q^2)\,\sigma+\zeta\,
  {\cal B}$  \\ [2ex] 
${\cal D}$ & $\displaystyle -2\,\frac{V}{M_B+m_{K^*}}\,\sigma $ & $2\, 
g\, \sigma $ \\[2ex]
${\cal E}$ & $\displaystyle \frac{A_0}{M_B+m_{K^*}}\, \Delta\, \sigma $ 
& $-f\, \sigma $ \\[1ex]
${\cal F}$ & $\displaystyle 2\,\frac{A_{+}}{M_B+m_{K^*}}\,W\cdot 
q\,\sigma-\zeta
   \,{\cal E} $ & $\displaystyle -2\,a_{+}\,W.q\,\sigma -\zeta\,{\cal E} 
   $\\ [2ex]
\end{tabular}      
\vskip 0.2truecm
\end{table}
\end{center}

\widetext
\begin{center}
\begin{table}
\squeezetable
\caption{The matrix element mod. squared for $$B$\to K\pi \ell^+\ell^-$.}
\label{tab2}
\begin{tabular}{ll}
$\left|{\cal M}
\right|^2=
\displaystyle\frac{\alpha^2 G_F^2}{2\pi^2} \displaystyle
\Bigl(
   \displaystyle 2\epsilon_{\mu\nu\rho\sigma}k^\mu K^\nu q^\rho Q^\sigma 
\Bigl(
    K\cdot Q\,{\rm Im}(\alpha _{L}\beta_{L}^* +
    \alpha _{R}\beta _{R}^*)+
    {\rm Im}(\rho _{L}\,\beta _{L}^* -\rho _{R}\,\beta _{R}^*)\displaystyle
 -\,k\cdot Q\,{\rm Im}(\rho _{L}\,\alpha _{L}^*+
   \rho _{R}\,\alpha _{R}^*) \Bigr)$&\\[2ex]$
  + 2\, {\rm Re}(\rho _{R}\,\alpha _{R}^* -\rho _{L}\,\alpha _{L}^*) 
  \left( -k\cdot K  \,q^2\,k\cdot Q  +
     k\cdot q\,k\cdot Q\,K\cdot q\right.\displaystyle 
     \left.+\, m_{K^*}^2\,q^2\,K\cdot Q -
     {k\cdot q}^2\,K\cdot Q \right) \,+
  2\,{\rm Re}(\rho _{L}\,\beta _{L}^*+\rho _{R}\,\beta _{R}^*)\,
 \left( k\cdot q K\cdot q \right.$&\\[2ex]$\left.- k\cdot K q^2-
     k\cdot Q K\cdot Q \right)+
     2\,{\rm Re}(\alpha _{L}\,\beta _{L}^* -\alpha _{R}\,\beta _{R}^*)
\left( K^2\,q^2\,k\cdot Q - k\cdot Q\,{K\cdot q}^2 -
     k\cdot K\,q^2\,K\cdot Q+
 k\cdot q\,K\cdot q\,K\cdot Q \right)+\,
  \left( \rho_L^2+\rho_R^2 \right)$&\\[2ex]$\,
\left( -m_{K^*}^2\,q^2 + k\cdot q^2-k\cdot Q^2 \right)+    
  \left( {{\beta_{L}}^2} +{{\beta_{R}}^2} \right)
 \left( -K^2\,q^2+{K\cdot q}^2 -
     {K\cdot Q}^2 \right)\displaystyle\,+
  \left( {\alpha_L}^2 + {\alpha_R}^2\right)\,
\left( -K^2\,q^2\,{k\cdot q}^2   +
     {k\cdot Q}^2\,{K\cdot q}^2\right.$&\\[2ex]$\left.+
    2\,k\cdot K\,q^2\,k\cdot Q\,K\cdot Q -
      m_{K^*}^2\,q^2\,{K\cdot Q}^2 +
     {k\cdot q}^2\,{K\cdot Q}^2-
    2\,k\cdot q\,k\cdot Q\,K\cdot q\,K\cdot Q  \right)                
\Bigr)\,$,\\[3ex]
$k\cdot K =m_K^2-m_\pi^2                      \,,\;\displaystyle
k\cdot q =\displaystyle\frac{(\Delta-q^2)}{2},\,\Delta=(M_B^2-m_{K^*}^2)        
\,,\;\displaystyle
k\cdot Q =  X M_B \cos\theta_e                 \,,\;\displaystyle  
X=\frac{(k\cdot q^2-q^2 m_{K^*}^2)^\frac{1}{2}}{M_B}         ,\,\;
\lambda_e=\sqrt{1-\frac{4\,m_e^2}{q^2}}
\,,$&\\$\displaystyle 
K\cdot Q =\lambda_K( k\cdot q \cos\theta_e \cos\theta_K -\sqrt{q^2}m_{K^*}
    \sin\theta_e \sin\theta_K \cos\varphi )
  +k\cdot q\,\zeta,\,\zeta=\displaystyle\frac{k\cdot K}{m_{K^*}^2}                  
  \,,\displaystyle   
K\cdot q =  \lambda_K X M_B \cos\theta_{K} + k\cdot q\,z   ,\; \displaystyle  
q\cdot Q=0                                                   
$&\\$\displaystyle
\lambda_K=\left(1-\displaystyle\frac{(m_K+m_\pi)^2}{m_{K^*}^2}
\right)^\frac{1}{2} 
   \left(1-\displaystyle\frac{(m_K-m_\pi)^2}{m_{K^*}^2}
   \right)^\frac{1}{2},\; 
   \displaystyle
\epsilon_{\mu\nu\rho\sigma} k^\mu K^\nu q^\rho Q^\sigma =-                              \displaystyle
X M_B \lambda_K \sqrt{q^2} m_{K^*}
   \sin\theta_e \sin\theta_K \sin\varphi                                
\;.$
\end{tabular}
\end{table}
\end{center}
\narrowtext
\noindent
\end{document}